# Beyond the band edge: Unveiling high-mobility hot carriers in a two-dimensional conjugated coordination polymer

Shuai Fu[1,2,12], Xing Huang[1,12], Guoquan Gao[3,12], Petko St. Petkov[4], Wenpei Gao[5], Jianjun Zhang[1], Lei Gao[2], Heng Zhang[2], Min Liu[2], Mike Hambsch[6], Wenjie Zhang[7], Jiaxu Zhang[1], Keming Li[3], Ute Kaiser[8], Stuart S. P. Parkin[7], Stefan C. B. Mannsfeld[6], Tong Zhu[3]*, Hai I. Wang[2,9]*, Zhiyong Wang[1,7]*, Renhao Dong[10,11]*, Xinliang Feng[1,7]*, Mischa Bonn[2]*

[1]Center for Advancing Electronics Dresden and Faculty of Chemistry and Food Chemistry, Technische Universität Dresden, Dresden, Germany

[2]Max Planck Institute for Polymer Research, Mainz, Germany

[3]Laser Micro/Nano Fabrication Laboratory, School of Mechanical Engineering, Beijing Institute of Technology, Beijing, China

[4]Faculty of Chemistry and Pharmacy, University of Sofia, Sofia, Bulgaria

[5]State Key Laboratory of Metal Matrix Composites, School of Materials Science and Engineering, Future Material Innovation Center, Zhangjiang Institute for Advanced Study, Shanghai Jiao Tong University, Shanghai, China

[6]Center for Advancing Electronics Dresden and Faculty of Electrical and Computer Engineering, TUD Dresden University of Technology, Germany

[7]Max Planck Institute of Microstructure Physics, Halle (Saale), Germany

[8]Central Facility for Materials Science Electron Microscopy, Universität Ulm, Ulm, Germany

[9]Nanophotonics, Debye Institute for Nanomaterials Science, Utrecht University, Utrecht, the Netherlands

[10]Department of Chemistry, The University of Hong Kong, Hong Kong 999077, China

[11]Materials Innovation Institute for Life Sciences and Energy (MILES), HKU-SIRI, Shenzhen 518048, China

[12]These authors contributed equally: Shuai Fu, Xing Huang, Guoquan Gao

*E-Mail: tongzhubit@bit.edu.cn; h.wang5@uu.nl; wang.zhiyong@tu-dresden.de; rhdong@hku.hk; xinliang.feng@tu-dresden.de; bonn@mpip-mainz.mpg.de

**Abstract**

Hot carriers, inheriting excess kinetic energy from high-energy photons, underpin numerous optoelectronic applications involving non-equilibrium transport processes. Current research on hot carriers has predominantly focused on inorganic materials, with little attention paid to organic-based systems due to their ultrafast energy relaxation and inefficient charge transport. Here, we overturn this paradigm by demonstrating highly mobile hot carriers in solution-processable, highly crystalline two-dimensional conjugated coordination polymer (2D *c*-CP) $Cu_3BHT$ (BHT = benzenehexathiol) films. Leveraging a suite of ultrafast spectroscopic and imaging techniques, we unravel the microscopic charge transport landscape in $Cu_3BHT$ films following non-equilibrium photoexcitation across temporal, spatial, and frequency domains, revealing two distinct high-mobility transport regimes. In the non-equilibrium transport regime, hot carriers achieve ultrahigh mobility of ~2,000 $cm^2$ $V^{-1}$ $s^{-1}$, traversing grain boundaries up to ~300 nm within a picosecond. In the quasi-equilibrium transport regime, free carriers exhibit Drude-type band-like transport with a remarkable mobility of ~400 $cm^2$ $V^{-1}$ $s^{-1}$ and an intrinsic diffusion length exceeding 1 μm. These findings establish 2D *c*-CPs as versatile platforms for exploring high-mobility non-equilibrium transport, unlocking new opportunities for organic-based hot carrier applications.

**Main**

Hot carriers refer to high-energy electrons and holes out of thermal equilibrium with the crystal lattice. They exhibit fascinating transport phenomena and distinctive physicochemical properties that have catalyzed innovations in various fields, such as photovoltaics, transistors, photodetectors, photocatalysis, and bolometers.[1-6] The possibilities of hot carriers have driven a surge in both computational and experimental studies over the past few decades.[7-15] Accumulating evidence has delineated a generalized three-stage fate for hot carriers: (i) generation, where electrical or optical excitations transiently populate high-energy electronic states in a non-equilibrium manner; (ii) thermalization, where carrier-carrier collisions transform the electronic system from a non-thermal distribution to a thermal distribution obeying Fermi-Dirac statistics at an elevated carrier temperature; and (iii) hot carrier relaxation, where inelastic carrier-phonon interactions lead to energy dissipation, establishing a quasi-thermal equilibrium between the electronic system and the lattice. Eventually, the lattice returns to its equilibrium state through thermal energy dissipation into the surrounding environment. In stark contrast to the growing understanding of hot carriers in inorganic or hybrid material systems, the study and application of hot carriers in organic compounds lag far behind. This gap primarily arises from the ultrafast energy relaxation process occurring on the femtosecond (fs) timescale in conventional organic compounds,[16,17] compounded by their limited charge transport capabilities due to pervasive dynamic disorder, strong Coulomb interactions between electron-hole pairs, and intense charge-vibration coupling.[8,18-21] As a result, organic compounds face significant challenges in achieving viability for organic-based hot carrier applications, which are actively sought after in the field. Overcoming these limitations and harnessing the potential of hot carriers in organic-based materials remains an important area of research, with the potential to unlock new possibilities in various technological applications.

The recent rise of synthetic organic two-dimensional (2D) crystals, particularly 2D conjugated coordination polymers (2D *c*-CPs) or 2D conjugated metal-organic frameworks (2D *c*-MOFs), may offer a pivotal opportunity to bridge this gap.[22] These 2D *c*-CPs, constructed by square-planar linkages and $\pi$ conjugated ligands, exhibit appealing electronic properties due to tunable intralayer *d*-$\pi$ conjugation and interlayer electronic coupling.[22-25] Recent advances in synthetic strategies allow precise control over stoichiometry, structural topology, layer orientation, and morphology, giving rise to a host of exquisite structures with diverse properties.[26-31] Two striking properties that stand out are their high electrical conductivity, up to $10^3$ S cm$^{-1}$,[32,33] on par with metals, and their low thermal conductivity below 1 W m$^{-1}$ k$^{-1}$,[34] akin to linear polymers (**Fig. 1a** and **Supplementary Tables 1−4**). These two seemingly contradictory characteristics — functioning as electrical conductors while being thermal insulators — bring 2D *c*-CPs to the forefront of thermoelectric and hot carrier applications. The thermoelectric behavior is reminiscent of the electron-crystal phonon-glass paradigm, which is crucial for achieving a high thermoelectric figure of merit.[35,36] In the context of hot carrier applications, 2D *c*-CPs offer the opportunity to engineer electronic and phononic properties at the molecular scale, enabling the creation of high-mobility charge transport channels while simultaneously decelerating hot carrier cooling processes.[24,37-39] Despite these promising prospects, to the best of our knowledge, no study has yet observed the signatures of hot carriers in 2D *c*-CPs, let alone evaluated their transport characteristics and discovered high-mobility non-equilibrium electronic states that are paramount for hot carrier applications.

Here, we combine time-resolved terahertz spectroscopy (TRTS), transient absorption spectroscopy (TAS), and transient absorption microscopy (TAM) to visualize the spatiotemporal evolution of non-equilibrium photoexcitation and track the excess kinetic energy-dependent charge transport properties in a model 2D *c*-CP system, namely $Cu_3BHT$ (BHT = benzenehexathiol) (**Fig. 1b**). Jointly, these experiments provide a comprehensive understanding of transport phenomena across temporal, spatial, and frequency domains. Our findings reveal that above-bandgap photoexcitation launches a transport cascade in highly crystalline $Cu_3BHT$ films. Hot carriers dominate the non-equilibrium transport regime, exhibiting exceptionally high charge mobility of ~2,000 $cm^2$ $V^{-1}$ $s^{-1}$ and migrating up to ~300 nm across grain boundaries. Owning to the low optical phonon energy and small electron-hole reduced effective mass, the cooling of these highly mobile hot carriers occurs on relatively long timescales, up to ~750 fs, comparable to state-of-the-art lead-halide perovskites known for their suitability for hot carrier applications (with cooling times typically ranging from 200 fs to a few ps).[40] The energy relaxation process is subject to a pronounced hot phonon bottleneck effect at high photoexcitation densities, likely originating from acoustic-optical phonon up-conversion. Upon entering the quasi-equilibrium transport regime, band-edge carriers exhibit band-like Drude-type free carrier transport with an impressive charge mobility of ~400 $cm^2$ $V^{-1}$ $s^{-1}$ and a remarkably long intrinsic diffusion length exceeding 1 μm.

**Results and discussion**

## Synthesis and characterization of Cu$_3$BHT films

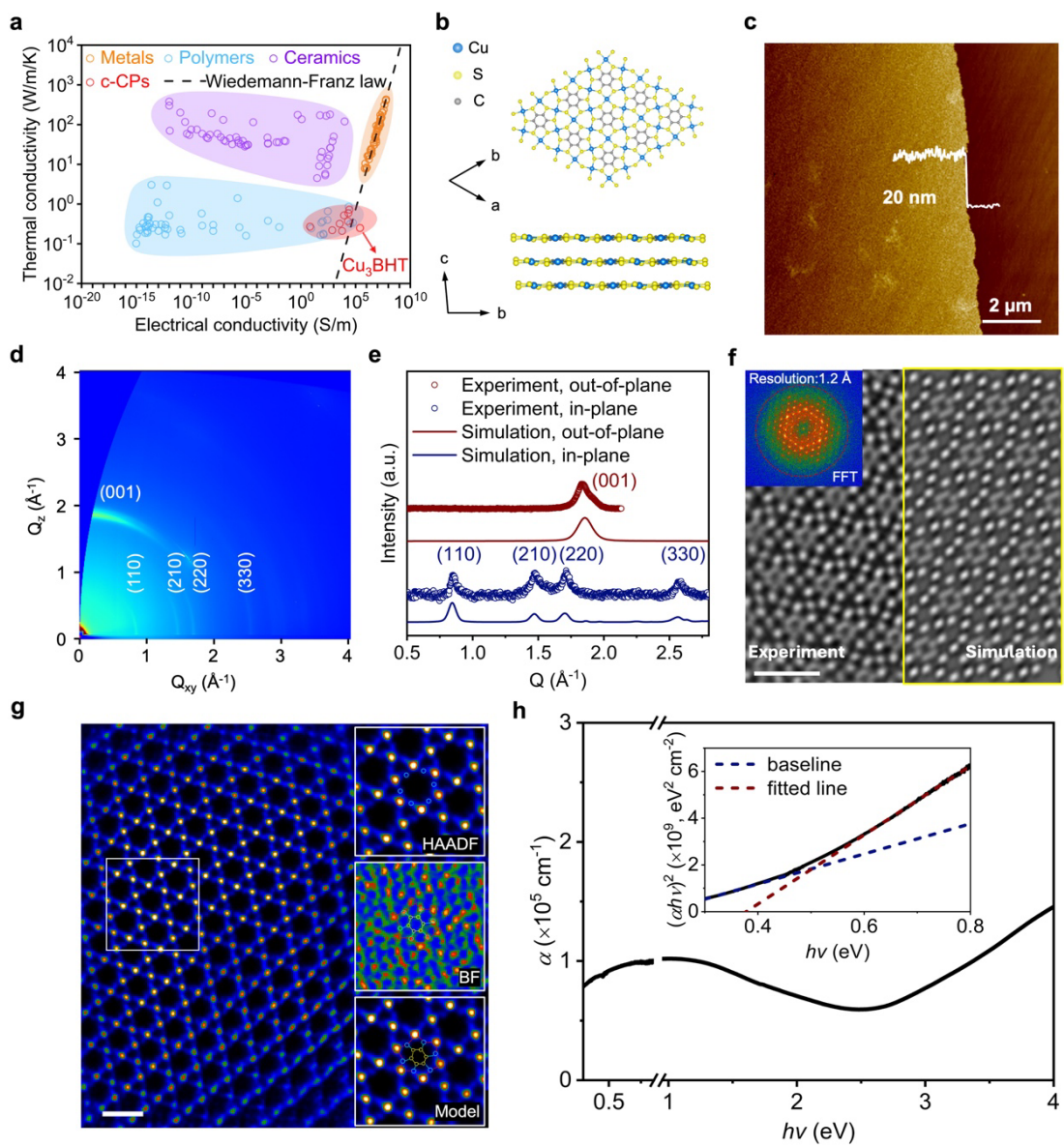

**Fig. 1 | Characterization of the Cu$_3$BHT films synthesized by liquid-liquid interfacial reaction. a**, Ashby plot illustrating thermal conductivity versus electrical conductivity for various material systems. **b**, Crystal structure of Cu$_3$BHT viewed along the stacking direction (top) and from the side (bottom). **c**, AFM image of Cu$_3$BHT film on a fused silica substrate. **d**, 2D GIWAXS image of Cu$_3$BHT. **e**, Experimental and calculated GIWAXS intensity profiles of Cu$_3$BHT, projected along the in-plane and out-of-plane directions. **f**, AC-HRTEM image and corresponding FFT pattern of Cu$_3$BHT (left) and its simulated image (right), denoised using Wiener filtering (scale bar: 1 nm). **g**, Atomic-resolution HAADF-STEM image of Cu$_3$BHT (scale bar: 1 nm). The three panels on the right display the HAADF, BF, and model images, respectively. **h**, Optical absorption spectrum of Cu$_3$BHT film. Insert: Data in the infrared range plotted using Tauc units.

Cu$_3$BHT films were synthesized via a liquid-liquid interfacial synthesis method following an optimized protocol derived from previous reports (see **Methods** for details).[32,33] Coordination polymerization between Cu$^{2+}$ and BHT ligands at the water/toluene interface enables the formation of a large-area Cu$_3$BHT film that uniformly covers the $1 \times 1$ cm$^2$ substrate with a thickness of ~20 nm (**Fig. 1c** and **Supplementary Fig. 1**). The Raman spectrum (**Supplementary Fig. 2**) of the Cu$_3$BHT film transferred onto a fused silica substrate displays low-energy optical phonon branches around 100 cm$^{-1}$ and intense bands in the ranges of 250−550 cm$^{-1}$ and 1000−1700 cm$^{-1}$, attributed to the vibration modes of Cu and S atoms, the interactions between C and S atoms, and the in-plane C stretching, respectively.[41] The absence of the −SH stretching vibration around 2500 cm$^{-1}$ confirms the effective coordination between Cu$^{2+}$ and −SH groups of BHT ligands.[42,43] To resolve the crystal structure of the as-synthesized Cu$_3$BHT thin film with atomic precision, we employed aberration-corrected high-resolution transmission electron microscopy (AC-HRTEM), high-angle annular dark-field scanning transmission electron microscopy (HAADF-STEM), and grazing-incidence wide-angle X-ray scattering (GIWAXS). AC-HRTEM imaging (**Fig. 1f**, left) reveals a highly ordered lattice with atomic resolution (~1.2 Å) and a lattice spacing of ~0.73 nm, in good agreement with the (100) plane of Cu$_3$BHT. The HAADF-STEM image (**Fig. 1g**) visualizes a high-symmetry, non-distorted Kagome lattice formed by Cu atoms, where each BHT unit connects to six neighboring BHT units via shared Cu atoms. Each Cu atom coordinates with four S atoms, forming a dense hexagonal *d-π* conjugated plane. These observations indicate that the as-synthesized Cu$_3$BHT thin film exhibits distinct stacking characteristics compared to the previously reported Cu$_3$BHT bulk crystals with an AB stacking mode.[44] The GIWAXS pattern (**Fig. 1d** and **1e**) shows a distinct arc at $Q_z$ = 1.85 Å$^{-1}$, corresponding to the (001) plane along the π-π stacking direction, indicating a preferential face-on orientation with an interlayer distance of 3.4 Å. Bragg diffraction peaks are observed at $Q_{xy}$ = 0.85 Å$^{-1}$, 1.47 Å$^{-1}$, 1.70 Å$^{-1}$, and 2.55 Å$^{-1}$, which can be indexed to the in-plane structure. By integrating the well-defined hexagonal structure revealed by AC-HRTEM with insights from the GIWAXS analysis, we proposed a triclinic lattice structure, featuring an slipped-AA stacking geometry and unit cell parameters of $a = b = 8.675$ Å, $c = 3.489$ Å, $\alpha = \beta = 99.94°$, and $\gamma = 60.12°$. The simulated AC-HRTEM image (**Fig. 1f**, right) and GIWAXS diffraction signals (**Supplementary Fig. 3**), based on this model, align well with the experimental results, further validating the proposed structure.

X-ray photoelectron spectroscopy (XPS) and extended X-ray absorption near-edge structure (XANES) measurements were carried out to elucidate the oxidation state of Cu. The high-resolution XPS spectrum of Cu 2$p$ (**Supplementary Fig. 4**) shows two asymmetric peaks at 952.9 and 932.9 eV, corresponding to Cu 2$p_{1/2}$ and Cu 2$p_{3/2}$, respectively, alongside satellite features indicative of $Cu^{2+}$. Deconvolution of the Cu 2$p$ regions reveals the coexistence of $Cu^+$ and $Cu^{2+}$ oxidation states. The XANES spectra (**Supplementary Fig. 5**) further corroborate the presence of mixed valence states of Cu, with the K-edge of $Cu_3BHT$ situated between those of $Cu_2O$ and $CuO$. The Cu K-edge Fourier-transformed extended X-ray absorption fine structure (EXAFS) $R$-space spectrum (**Supplementary Fig. 6**) showcases a prominent peak at 1.87 Å, attributed to tetra-coordinated Cu-S bonds, along with minor peaks associated with higher shell Cu-C and Cu-Cu scattering. Fitting analysis reveals an average Cu coordination number of 3.22 (± 0.32) and a Cu-S bond length of 2.29 Å, consistent with the square planar coordination geometry observed above. Collectively, the XPS and XANES results indicate a fractional Cu oxidation state, which can be attributed to an intramolecular pseudo-redox mechanism between $Cu^+/Cu^{2+}$ and BHT ligands.[45,46]

Given the potential influence of minor changes in metal valence states on electronic properties,[47,48] we further investigated the electronic band structures of $Cu_3BHT$ under different Cu valence state configurations using density functional theory (DFT) calculations (see **Supplementary Information** for details). These calculations reveal that gradually reducing the $Cu^{2+}/Cu^+$-ratio from 1 to 0 induces a surprising trend of bandgap opening, although the band dispersion is largely preserved (**Supplementary Fig. 7**). This suggests that the Cu valence state can sever as an effective knob to tune the optical and electrical properties of $Cu_3BHT$. Optical absorption measurements of the as-prepared $Cu_3BHT$ film reveal a broadband absorption feature spanning the UV-visible-near-infrared range (**Fig. 1h**). Plotting the absorption data in Tauc units reveals an absorption edge at ~0.5 eV (**Fig. 1h**, insert), which we tentatively assign to optical transitions associated with the in-plane energy gap. This gapped nature is further validated by ultraviolet photoelectron spectroscopy (UPS), which demonstrates an energy separation of ~0.2 eV between the valence band edge and the Fermi level (**Supplementary Fig. 8**). Variable-temperature conductivity from 2 K to 300 K using a four-terminal configuration (**Supplementary Fig. 9**) reveal a negative temperature coefficient of conductivity, supporting the semiconducting behavior inferred from optical absorption and UPS measurements. Note that while the room-temperature electrical conductivity (~48 S cm$^{-1}$) of the synthesized Cu$_3$BHT film surpasses most state-of-the-art 2D $c$-CPs,[26] it is substantially lower than that of metallic phase Cu$_3$BHT bulk crystals featuring an AB stacking mode and interlayer Cu-S covalent bonds (up to 2,500 S cm$^{-1}$),[33,44] further highlighting the semiconducting nature of the synthesized Cu$_3$BHT thin film.

**Observation of highly mobile hot carriers and hot phonon bottleneck in Cu$_3$BHT**

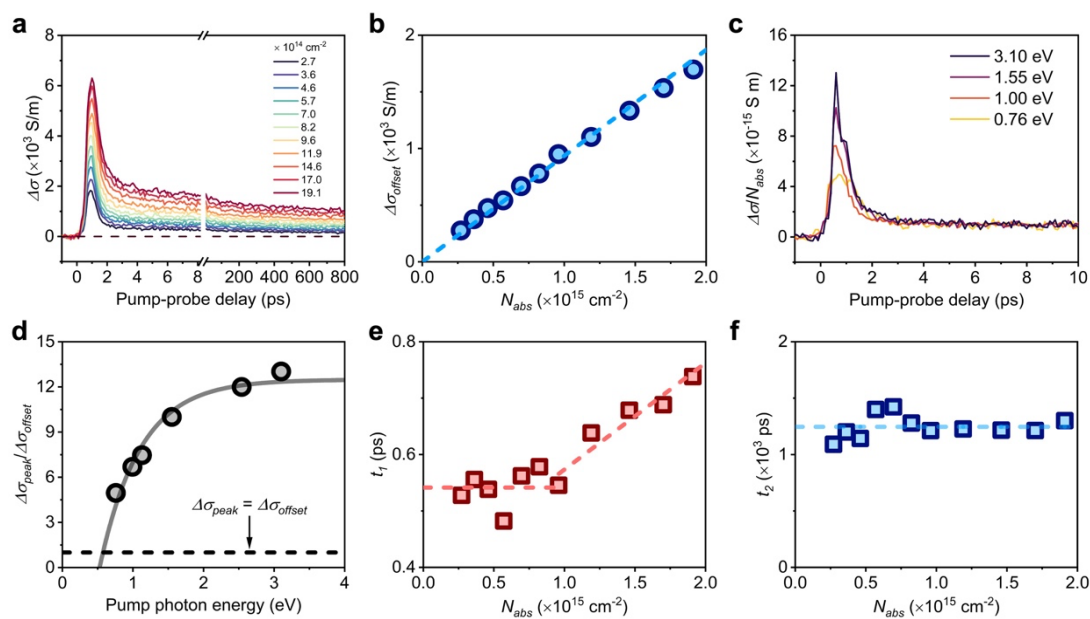

**Fig. 2 | Non-equilibrium photoexcitation cascades and hot phonon bottleneck in Cu$_3$BHT. a,** THz photoconductivity dynamics following 1.55 eV excitation at different absorbed photon fluence ($N_{abs}$). **b,** $\Delta\sigma_{offset}$, defined as the average THz photoconductivity between 6 and 8 ps at different $N_{abs}$. The dashed line represents a linear fit through the origin of the coordinate system. **c,** Pump photon energy ($h\nu$)-dependent THz photoconductivity dynamics normalized by absorbed photon number

$N_{abs}$. **d,** Ratio of the maximum photoconductivity ($\Delta\sigma_{peak}$) to the average THz photoconductivity between 6 and 8 ps ($\Delta\sigma_{offset}$) as a function of photon energy $h\nu$. The solid grey line represents the extrapolation used to estimate the critical photon energy value at which the ratio equals 1 (dashed line). **e,** Time constants associated with hot carrier cooling, $t_1$, and **f,** charge recombination, $t_2$, versus $N_{abs}$ inferred from the data in panel **a**. The dashed lines are guides to the eye.

We employ time-resolved terahertz spectroscopy (TRTS) as a contact-free, non-invasive approach to explore the dynamics and transport properties of photogenerated charge carriers in Cu$_3$BHT films. In TRTS measurements, an ultrashort pump pulse (~50 fs duration) with tunable photon energy photoinjects hot carriers into the sample via above-gap excitations. Subsequently, a time-delayed, single-cycle THz electromagnetic pulse (~1 ps duration) propagates through the sample, driving the photogenerated carriers over short distances (typically sub-ten to tens of nm), affording insights into their intra-crystal charge transport properties.[49] Details on TRTS and photoconductivity analysis are provided in the **Methods** section. **Fig. 2a** shows the pump fluence-dependent THz photoconductivity dynamics following 1.55 eV excitation. The non-resonant excitation induces a rapid sub-ps rise in photoconductivity due to the quasi-instantaneous generation of mobile carriers. The positive THz photoconductivity, combined with the highly crystalline structure revealed by AC-HRTEM and HAADF-STEM, further corroborates the intrinsic semiconducting nature of the synthesized Cu$_3$BHT films.[37] This is followed by a photoconductivity decay characterized by a fast-decay component within ~1 ps and a long-lived component persisting for over ~1 ns. Since the photoconductivity $\Delta\sigma(t)$ is determined by the product of photogenerated carrier density ($n$), elementary charge ($e$), and electron-hole sum mobility ($\mu$) via $\Delta\sigma(t) = n \cdot e \cdot \mu$, two possible scenarios can account for the fast-decay component: (i) hot carriers rapidly lose their excess kinetic energy and populate band-edge states within the instrument's time resolution (< ~50 fs), with the fast-decay component reflecting a reduction in band-edge carrier density due to fast charge localization or trapping ($n$ decays quickly after photoexcitation);[50,51] (ii) hot carriers possess much higher charge mobility than band-edge carriers and energy relaxation occurs on a timescale long relative to the instrument's time resolution (> ~50 fs), with the fast-decay component signifying a decrease in charge mobility as the non-equilibrium electronic system relaxes to the band edge ($\mu$ drops quickly following photoexcitation and hot carrier relaxation).[52] For clarity, we define the maximum photoconductivity as $\Delta\sigma_{peak}$ and the average photoconductivity value between 6 to 8 ps as $\Delta\sigma_{offset}$. A key criterion for distinguishing these two scenarios lies in the dependence of $\Delta\sigma_{offset}$ on the absorbed photon density ($N_{abs}$). As shown in **Fig. 2b**, the linear dependence of $\Delta\sigma_{offset}$ on $N_{abs}$ (i) excludes the dominant role of defect trapping scenario, which would otherwise exhibit a characteristic super-linear dependence[50,51], (ii) demonstrates that band-edge carrier density increases linearly with $N_{abs}$, without hitting the threshold for absorption saturation or non-radiative Auger relaxation, and (iii) indicates that $\mu$ of band-edge carriers remains constant, with negligible carrier-carrier interactions within the investigated $N_{abs}$ range.

To further validate that cooling of highly mobile hot carriers drives the fast-decay component, we varied the excitation photon-energy ($h\nu$), and performed TRTS measurements in the low $N_{abs}$-limit (<1×10$^{14}$ cm$^{-2}$) to minimize the scattering between hot carriers and hot phonons (**Supplementary Fig. 10**). **Fig. 2c** compares the photoconductivity response normalized to $N_{abs}$ ($\Delta\sigma/N_{abs}$), following optical excitation at different $h\nu$. Within the first 10 ps time where $n$ can be considered constant, $\Delta\sigma/N_{abs}$ provides a direct view of the temporal evolution of $\mu$. Qualitatively, we find that higher $h\nu$ results in an increased transient $\mu$ within the first ps, after which $\mu$ drops substantially until reaching the same value regardless of $h\nu$. Extrapolation of $\Delta\sigma_{peak}/\Delta\sigma_{offset}$ to $h\nu$ (**Fig. 2d**, see **Methods** for details) suggests that the critical photon energy corresponding to resonant excitation is ~560 ± 50 meV, in good agreement with the absorption edge revealed by Tauc analysis. These results indicate that hot carriers exhibit much higher $\mu$ than band-edge carriers in Cu$_3$BHT. Accordingly, the subsequent long-lived component can be assigned to the recombination process involving band-edge electrons and holes in the quasi-equilibrium regime.

By extracting the time constants associated with hot carrier cooling ($t_1$) and band-edge carrier recombination ($t_2$) at different $N_{abs}$ using a bi-exponential decay function, we find that $t_1$ exhibits a stepwise variation with $N_{abs}$, whereas $t_2$ remains unchanged, irrespective of $N_{abs}$ (**Fig. 2e** and **Supplementary Fig. 11**). Specifically, $t_1$ stays at ~500 fs within experimental accuracy when $N_{abs}$ is below $10^{15}$ cm$^{-2}$. However, once $N_{abs}$ surpasses this threshold, $t_1$ gradually increases to ~750 fs as $N_{abs}$ rises, a hallmark of the hot phonon bottleneck.[53] On the other hand, the invariance of $t_2$ ~1.2 ns across the studied $N_{abs}$ range implies that trap-assisted recombination is likely the dominant recombination mechanism. The observed hot phonon bottleneck can be understood by noting the phonon dispersion in Cu$_3$BHT,[41] where low-energy optical phonon branches around ~100 cm$^{-1}$ intersect with acoustic phonon branches. The low optical phonon energy and small electron-hole reduced effective mass favor reduced energy dissipation rates.[13] This results in a relatively long hot carrier lifetime in Cu$_3$BHT, superior to conventional organic compounds (typically below 100 fs)[16,17] and comparable to inorganic and hybrid perovskite materials (ranging from 200 fs to a few ps)[40] that are promising for hot carrier applications. Furthermore, the considerable phononic overlap, combined with the limited phonon propagation due to the low thermal conductivity of Cu$_3$BHT, may facilitate the up-conversion of acoustic phonons to optical phonons, which is responsible for the observed hot phonon bottleneck. A similar phenomenon has been reported in the perovskite formamidinium lead iodide, where organic cations introduce overlapping phonon branches, enhancing anharmonic phonon-phonon scattering and enabling a hot phonon bottleneck via up-transitions of low-energy phonon modes.[54] We note that the mechanism underlying the hot phonon bottleneck in Cu$_3$BHT differs markedly from the widely used phononic bandgap opening strategy. In the phononic bandgap opening strategy, elements or units with large atomic mass differences are used to create a phononic bandgap larger than twice the highest acoustic phonon energy, thereby suppressing the Klemens-decay pathway (i.e., a longitudinal optical phonon decays into two counter-propagating longitudinal acoustic phonons).

**Crossover from non-equilibrium to quasi-equilibrium transport regime**

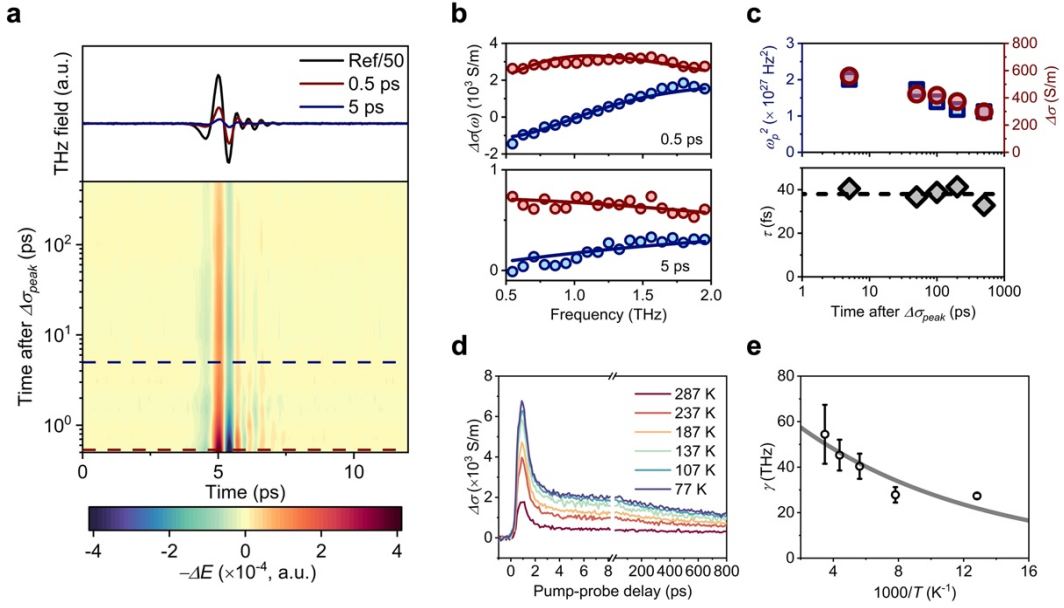

**Fig. 3 | Spectral signature of the transition from hot charge to band-edge carrier. a,** The top panel shows the time-resolved THz electric field transmitted through unexcited $Cu_3BHT$ and the pump-induced time-resolved THz electric field changes at representative times after $\Delta\sigma_{peak}$. The bottom panel displays a pseudo-color plot of the pump-induced time-resolved THz electric field changes at different times after $\Delta\sigma_{peak}$. The results are collected under 1.55 eV photoexcitation at $N_{abs} = 0.6 \times 10^{15}$ cm$^{-2}$ at room temperature. **b,** Frequency-resolved complex THz photoconductivity measured at 0.5 ps (top) and 5 ps (bottom) after $\Delta\sigma_{peak}$. The red and blue solid lines are the Drude-Smith (top) and Drude (bottom) fits, describing the real and imaginary components of the complex THz photoconductivity. **c,** The top panel shows squared plasma frequencies (blue squares, left $y$-axis) inferred from the Drude fits and the photoconductivity (red circles, right $y$-axis) at different times after $\Delta\sigma_{peak}$. The bottom panel shows charge scattering times inferred from the Drude fits at different times after $\Delta\sigma_{peak}$. **d,** Temperature-dependent THz photoconductivity dynamics. The results are collected under 1.55 eV photoexcitation at $N_{abs} = 0.4 \times 10^{15}$ cm$^{-2}$. **e,** Charge scattering rates inferred from the Drude fits at different temperatures. Error bars represent standard deviations obtained from the Drude fits. The solid line is a fit according to the Arrhenius relation.

To gain deeper insight into charge transport properties across different transport regimes, we track time-varying THz waveforms at various time delays after reaching $\varDelta\sigma_{peak}$ (**Fig. 3a**), from which we can infer the temporal evolution of frequency-resolved complex THz photoconductivity $\varDelta\sigma(\omega)$. We observe that $\varDelta\sigma(\omega)$ exhibits distinct frequency dispersions in the non-equilibrium and quasi-equilibrium transport regimes (**Fig. 3b**). Specifically, in the non-equilibrium transport regime, $\varDelta\sigma(\omega)$ shows suppressed real photoconductivity and negative imaginary photoconductivity at low frequencies. These spectral features reflect charge transport described by the phenomenological Drude-Smith model, where charge localization induced by backscattering events hinders long-range charge migration. The backscattering probability is quantified by a parameter $c$, which ranges from 0 (representing isotropic scattering, equivalent to the Drude model) to −1 (indicating complete backscattering) (see **Methods** for details).[55,56] In contrast, in the quasi-equilibrium transport regime, $\varDelta\sigma(\omega)$ shows positive real and imaginary components that converge with increasing frequency, indicative of delocalized free carrier transport, as described by the Drude model:

$$\Delta\sigma(\omega) = \frac{ne^2\tau}{m^*(1-i\omega\tau)}, \text{ with } n = \frac{\omega_p^2 m^* \varepsilon_0}{e^2}$$

where $\tau$, $m^*$, $\omega_p$, and $\varepsilon_0$ represent the momentum-averaged charge scattering time, electron-hole reduced effective mass, plasma frequency, and vacuum permittivity, respectively. The evolution of frequency dispersion during the transition from hot carriers to band-edge carriers can be understood as follows: hot carriers, with excess kinetic energy, can more easily navigate the fluctuating energy landscape and travel relatively long distances exceeding the grain size (~100 nm), thereby encountering a higher probability of backscattering (e.g., at grain boundaries) than band-edge carriers. Fitting $\varDelta\sigma(\omega)$ in the quasi-equilibrium transport regime with the Drude model yields $\tau$ of ~41 ± 3 fs. Using $m^* = 0.187$ $m_0$ from DFT calculations, $\mu$ of band-edge carriers in $Cu_3BHT$ is estimated to be 405 ± 30 $cm^2$ $V^{-1}$ $s^{-1}$ in the DC limit, following $\mu = e\tau/m^*$. By knowing the carrier lifetime $t_2$ and mobility, we estimate the intrinsic diffusion length of band-edge carriers to be ~1100 ± 300 nm (see **Methods** for details). By further considering the photoconductivity ratio between the non-equilibrium and quasi-equilibrium states under the photoexcitation conduction used, $\mu$ of hot carriers generated with $h\nu$ = 1.55 eV at $\Delta\sigma_{peak}$ is inferred to be approximately 2,000 cm$^2$ V$^{-1}$ s$^{-1}$. To the best of our knowledge, these values set new records for both mobility and diffusion length in 2D *c*-CPs as well as in organic materials. The Drude fits to $\Delta\sigma(\omega)$ at different time delays in the quasi-equilibrium transport regime reveal the temporal evolution of microscopic parameters (e.g., $n$ and $\tau$) related to charge transport during recombination. As shown in **Fig. 3c**, the inferred squared plasma frequency (proportional to $n$) follows the same trend as photoconductivity over time, indicating that the decrease in $n$ drives the photoconductivity decay, as a result of (likely trap-assisted) recombination. Meanwhile, $\tau$ remains largely unchanged in the quasi-equilibrium regime. This is consistent with a delocalized charge transport picture of a dilute free electron gas free from strong carrier-carrier interactions.[57]

To further explore the underlying charge transport mechanism, we perform temperature-dependent photoconductivity measurements from 287 K to 77 K. **Fig. 3d** shows the resulting photoconductivity dynamics, where both $\Delta\sigma_{peak}$ and $\Delta\sigma_{offset}$ increase substantially with decreasing temperature. Temperature-dependent $\Delta\sigma(\omega)$ measured in the quasi-equilibrium transport regime retains the spectral features of Drude-type transport (**Supplementary Fig. 12**). As shown in **Fig. 3e**, the inferred scattering rate $\gamma = 1/\tau$ increases from ~30 THz at 78 K to ~50 THz at 287 K, indicating that freezing out phonons or vibrational modes effectively reduces charge scattering. The positive temperature coefficient of $\gamma$ is a hallmark of band-like transport through extended electronic states. The temperature-$\gamma$ relationship reveals a critical energy value of 8 ± 2 meV (65 ± 15 cm$^{-1}$) following the Arrhenius relation (see **Methods** for details), pointing to a phonon mode of approximately that frequency involved in charge scattering.

## Spatiotemporal and energetic evolution of non-equilibrium photoexcitation

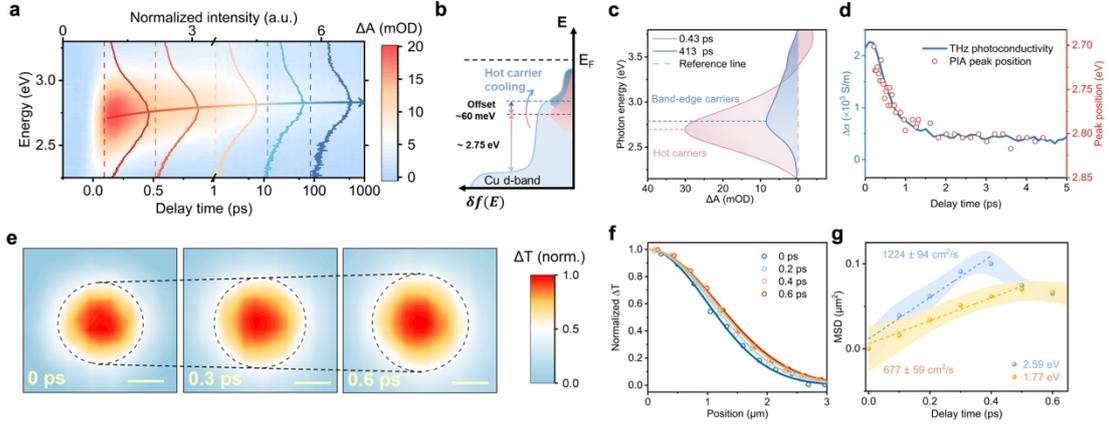

**Fig. 4 | Temporal and energetic evolution of hot carrier cooling and visualization of hot carrier transport in real space. a,** 2D pseudo-color TA spectrum as a function of probe photon energy and delay time under 1.77 eV photoexcitation at $N_{abs} = 1.1 \times 10^{15}$ cm$^{-2}$. The colored solid lines represent the corresponding normalized TA spectra at representative pump-probe delay times (i.e., 0.15, 0.5, 1, 10, and 100 ps). **b,** Schematic diagram of hot carrier cooling as the origin of the blue shift. **c,** Global fitting analysis deconvolutes the respective spectral contributions from hot and band-edge carriers. **d,** Comparison of THz photoconductivity decay and TA signal peak energy. **e,** Representative TAM images at different pump-probe delay times under 1.77 eV photoexcitation at $N_{abs} = 0.7 \times 10^{15}$ cm$^{-2}$ (scale bar: 1 μm). **f,** Spatial profiles (dots) and Gaussian fits (curves) of photogenerated charge carriers at different pump-probe delay times. **g,** Determination of hot carrier diffusion coefficients for two $h\nu$ at $N_{abs} = 0.7 \times 10^{15}$ cm$^{-2}$. The data are presented as variance with standard error. The dashed lines are linear fits used to extract the diffusion constants under different photoexcitation conditions.

Next, we explore photogenerated carriers' spatiotemporal and energetic evolution across different charge transport regimes using transient absorption spectroscopy (TAS) and transient absorption microscopy (TAM). Experimental details and analytical protocols for TAS and TAM are provided in the **Methods**. **Fig. 4a** presents a 2D pseudo-color map of Cu$_3$BHT, along with normalized cross-section TA spectra at representative pump-probe delay times under 1.77 eV photoexcitation. The spectra exhibit a broad photoinduced absorption (PIA) signal spanning the 2.4 to 3.0 eV range, with a noticeable blue shift of ~60 meV within the first few ps (**Fig. 4a** and **Supplementary Fig. 13**). Similar spectral signatures are also observed at higher $h\nu$ excitations (**Supplementary Fig. 14**). We find that the temporal evolution of the PIA band closely mirrors the THz photoconductivity dynamics, suggesting that the observed spectral blue shift is associated with hot carrier cooling.[14,58] This interpretation is further supported by the calculated electronic band structure and density of states, where transitions from the Cu *d*-band to the valence band edge can rationalize the observed PIA signal (**Fig. 4b** and **Supplementary Fig. 7**). To consolidate this, we perform a global fitting analysis of the TA spectra, identifying two distinct spectral components centered at 2.69 eV and 2.79 eV, respectively (**Fig. 4c**). Notably, the decay of the low-energy component coincides with the rise of the high-energy component on the ps timescale, while the high-energy component persists into the ns timescale (**Supplementary Fig. 15**). Given that their dynamic signatures correspond closely to those of hot carriers and band-edge carriers elucidated by TRTS, we assign the low-energy and high-energy components to hot carriers and band-edge carriers, respectively. **Fig. 4d** shows that the peak energy shift signifying hot carrier relaxation indeed evolves remarkably in tandem with the photoconductivity decay, substantiating the blue shift as a spectroscopic indicator of hot carrier cooling. On this basis, we assess the carrier temperature evolution by modeling the low-energy tail of the PIA signal as the high-energy tail of thermalized hot carriers, following the Fermi-Dirac distribution (**Supplementary Fig. 16**). The comparative analysis yields a consistent trend between the extracted carrier temperature and the THz photoconductivity, providing additional evidence for the high-mobility nature of hot carriers.

To directly visualize carrier migration in real space, we employ TAM with high temporal resolution and sub-diffraction-limit spatial precision. The time-evolving spatial distribution of the photoexcited population is quantitatively characterized by the mean-square displacement (MSD = $s_t^2$ - $s_0^2$), where $s$ is the Gaussian width of the population profile at a delay time $t$. Representative false-color 2D images of the signal intensity across spatial locations are shown in **Fig. 4e**. In the TAM experiments, both the pump and probe beams approach the diffraction limit, which is considerably larger than the average grain size. As a result, the spatiotemporal profiles capture contributions from both intra-crystal and inter-crystal charge transport. During the first ~0.6 ps, photogenerated carriers undergo ultrafast expansion, as depicted in the 2D image cross-sections in **Fig. 4f** and the extracted spatiotemporal profiles in **Fig. 4g**. By increasing $h\nu$ from 1.77 eV to 2.59 eV at constant $N_{abs}$ below the $10^{15}$ cm$^{-2}$ threshold of **Fig 2e**, we observe an increase in the diffusion coefficient ($D$) from 677 ± 59 cm$^2$ s$^{-1}$ to 1224 ± 94 cm$^2$ s$^{-1}$. The ultrafast diffusion timescale, exceptional charge transport properties, and enhanced charge transport properties at higher $h\nu$ are all consistent with the hot carrier characteristics identified by TRTS (**Fig. 2c**). Note that $D$ of hot carriers observed in the synthesized Cu$_3$BHT film is superior that of hot carriers reported in hybrid perovskites and metal films.[10,59,60] Together, these findings demonstrate from two complementary perspectives that the excess kinetic energy of hot carriers significantly boosts charge transport properties. Furthermore, under the same 1.77 eV excitation, $D$ of hot carriers decreases monotonously as $N_{abs}$ increases (**Supplementary Fig. 17**), which parallels the reduction in $\mu$ of hot carriers with rising $N_{abs}$ (**Supplementary Fig. 10**), likely due to the more pronounced scattering at elevated hot phonon populations. Depending on the photoexcitation conditions, the hot carrier propagation length ($l_t$) can be estimated to range from 200 to 320 nm, following $l_t = \sqrt{s_t^2 - s_0^2}$.[61] These long propagation lengths, far exceeding the average grain size, indicate the cross-boundary transport behavior of hot carriers. The ultrafast diffusion regime is followed by a contracting phase, which can tentatively be attributed to a consequence of a hot phonon bottleneck, as further justified by $N_{abs}$-dependent and $h\nu$-dependent measurements (**Supplementary Fig. 18**). This also rationalizes the earlier onset of the contraction phase at higher $h\nu$ or $N_{abs}$ (**Fig. 4g** and **Supplementary Fig. 10**).

   On a timescale of tens of ps, we identify a slower diffusion phase that differs from the initial ultrafast diffusion seen with highly mobile hot carriers, which we attribute to the quasi-equilibrium transport of band-edge carriers (**Supplementary Fig. 17**). Following 1.77 eV photoexcitation, $D$ of band-edge carriers ranges from 1.1 ± 0.2 cm$^2$ s$^{-1}$ to 1.9 ± 0.3 cm$^2$ s$^{-1}$ (corresponding to ambipolar mobility ($\mu_a$) from 42 ± 8 cm$^2$ V$^{-1}$ s$^{-1}$ to 72 ± 12 cm$^2$ V$^{-1}$ s$^{-1}$), with a declining trend as $N_{abs}$ increases. The relatively low mobility and its distinct dependence on $N_{abs}$, compared to those inferred from TRTS, can be reconciled by the different transport length scales each technique probes: TRTS characterizes local, intrinsic, intra-crystal transport with minimal influence from grain boundaries, featuring constant mobility when carrier-carrier interactions are negligible; TAM encompasses both intra-crystal and inter-crystal transport, exhibiting lower mobility that increases with elevated equilibrated electron-lattice temperatures due to more favorable thermally activated hopping transport across grain boundaries.[62]

**Outlook**

In this work, we employ complementary ultrafast spectroscopic and imaging techniques to elucidate how energy relaxation couples with charge transport across temporal, spatial, and frequency domains in 2D *c*-CPs. We reveal that non-resonant photoexcitation initiates a cascade transport process in 2D *c*-CPs: the non-equilibrium transport regime, governed by hot carriers, exhibits cross-boundary transport behavior with an ultrahigh $\mu$ of ~2,000 $cm^2$ $V^{-1}$ $s^{-1}$; the quasi-equilibrium transport regime, dominated by band-edge carriers, features Drude-type free carrier transport with an exceptional $\mu$ of ~400 $cm^2$ $V^{-1}$ $s^{-1}$ and an intrinsic diffusion length exceeding 1 μm. The first demonstration of high-mobility non-equilibrium states in 2D *c*-CPs opens up new possibilities for designing efficient charge transport pathways and advancing novel organic-based (opto-)electronic applications involving non-equilibrium transport, such as hot-electron transistors, hot carrier photovoltaics, and plasmonic photocatalysis. This premise is further reinforced by the extensive chemical and structural tunability of 2D *c*-CPs, along with recent breakthroughs in the fabrication of large crystals and ultrasmooth films of 2D *c*-CPs for device integration.[25,43] Overall, these findings only scratch the surface, with many opportunities to be explored through metal substitution and diversification, ligand design, and guest molecule interactions. Realizing this potential will require further interdisciplinary efforts in chemical design, theoretical calculations, spectroscopic studies, and application development.

**Methods**
**Synthesis of Cu$_3$BHT films**

Referring to previous reports[32,33], $Cu_3BHT$ films are synthesized via an interfacial reaction between two immiscible liquid media, namely, $CuSO_4/H_2O$ and BHT/toluene. First, the substrate of interest is placed into an empty beaker prior to synthesis. Then, 30 mL of $CuSO_4$ aqueous solution (0.5 mg mL$^{-1}$) and 30 ml of toluene are injected into the beaker in sequence, serving as the $Cu^{2+}$ source and buffer layer, respectively. After forming a stable liquid-liquid interface, 2 mL of BHT in toluene solution (0.1 mg mL$^{-1}$) is gently injected into the buffer layer to initiate the coordination polymerization between $Cu^{2+}$ and BHT at the interface. The formation of a dark-colored film at the liquid-liquid interface can be observed with the naked eye. After film formation, the liquid is gently removed using a syringe, allowing the formed film to settle naturally onto the target substrate. The obtained film is washed sequentially with methanol and acetone to remove potential impurities. Finally, the film is dried overnight under ambient conditions.

**Time-resolved terahertz spectroscopy**

Time-resolved terahertz spectroscopy was employed to track the time- and frequency-resolved photoconductivity of $Cu_3BHT$ films. The setup was powered by a Ti:sapphire mode-locked regenerative amplifier, which delivered ultrashort ~50 fs laser pulses centered at 1.55 eV, with a repetition rate of 1 kHz. Optical excitations at 1.55 eV and 3.10 eV were achieved by directly using a branch of the fundamental 1.55 eV laser pulse and by frequency doubling it with the aid of a $\beta$–$BiB_3O_6$ crystal, respectively. Optical excitations at other photon energies were obtained using a commercial optical parametric amplifier (Light Conversion). Single-cycle THz radiation with a duration of ~1 ps was generated and detected using a pair of 1 mm thick (110)–oriented ZnTe crystals through optical rectification and free-space electro-optic sampling, respectively. Measurements at room temperature were performed in transmission mode in a dry $N_2$-purged environment, while temperature-dependent measurements were carried out by placing the $Cu_3BHT$ film in a cryostat under vacuum conditions (pressure below $1 \times 10^{-4}$ mbar). The time-resolved photoconductivity was measured by fixing the sampling beam to the peak of the transient THz electric field and recoding the pump-induced signal intensity changes while varying the relative time delay between the pump pulse and the THz probe. The frequency-resolved complex THz photoconductivity $\Delta\sigma(\omega)$ was accessed by recording the time-varying THz profiles transmitted through the Cu₃BHT film with and without optical excitation ($E'(t)$ and $E(t)$), applying Fourier transform ($E'(\omega)$ and $E(\omega)$), and adopting the thin-film approximation:

$$\Delta\sigma(\omega) = -\frac{n_1 + n_2}{Z_0 l}\left(\frac{E'(\omega) - E(\omega)}{E(\omega)}\right)$$

where $Z_0 = 377\ \Omega$ is the impedance of free space, $n_1$ and $n_2$ are the refractive indices of the media before and after the Cu₃BHT film, and $l$ is the Cu₃BHT film thickness, respectively. To estimate the critical photon energy corresponding to resonant excitation ($E_g$), we fit $\frac{\Delta\sigma_{peak}}{\Delta\sigma_{offset}}$ to $h\nu$ using the following equation:

$$\frac{\Delta\sigma_{peak}}{\Delta\sigma_{offset}} = A\left(1 - e^{-\frac{h\nu - E_g}{E_g}}\right) + 1$$

where $A$ is the pre-factor. When $h\nu$ is equal to $E_g$, $\frac{\Delta\sigma_{peak}}{\Delta\sigma_{offset}} = 1$, reflecting the long-lived nature of band-edge carriers. When $h\nu$ is much larger than 1, $\frac{\Delta\sigma_{peak}}{\Delta\sigma_{offset}}$ tends to converge to a finite value, consistent with the fact that the electronic temperature gradually approaches saturation with increasing input energy.

The Drude-Smith model describing spatially confined charge transport of hot carriers reads:

$$\Delta\sigma(\omega) = \frac{\omega_p^2 \varepsilon_0 \tau_{DS}}{1 - i\omega\tau_{DS}}\left(1 + \frac{c}{1 - i\omega\tau_{DS}}\right)$$

where $\tau_{DS}$ is the Drude-Smith scattering time, and $c$ is the backscattering probability ranging from 0 (isotropic scattering) to −1 (completely backscattering).

The intrinsic diffusion length is calculated from $\mu$ and $t_2$ as follows:

$$L = \sqrt{\frac{\mu k_B T t_2}{e}}$$

where $k_B$ is the Boltzmann constant, $T$ is the temperature.

We estimate the activation energy ($E_a$) from the temperature-$\gamma$ relationship using the as Arrhenius relation:

$$\gamma = Be^{-(E_a/k_BT)}$$

where $B$ is the pre-factor.

**Transient absorption spectroscopy**

The femtosecond transient absorption setup utilized a regenerative-amplified Ti:sapphire laser system from Coherent and a Helios pump-probe system from Ultrafast Systems. The laser system delivered pulses with a central photon energy of 1.55 eV, a pulse duration of 25 fs, and a repetition rate of 1 kHz. The output beam from the amplifier was split into two branches: one beam passed through an optical parametric amplifier (TOPAS-C) to produce pump laser pulses with tunable photon energy, while the other beam was focused on a sapphire crystal to create a white-light continuum (WLC). The resulting WLC was split into a probe beam and a reference beam. The pump and probe pulses were precisely overlapped both spatially and temporally on the sample. A motorized optical delay line was utilized to adjust the pump-probe delay. The pump pulses were chopped by a mechanical chopper operating at 500 Hz, and the absorbance changes with and without the pump pulse were calculated.

**Transient absorption microscopy**

The output of a high-repetition-rate amplifier (PH1-20, Light Conversion, 800 kHz, 1030 nm) served as the input to an optical parametric amplifier (OPA, TOPAS-Twins, Light Conversion) with two independent outputs: one providing the pump beam and the other supplying the probe beam. Both the pump and probe beams were spatially filtered. An acousto-optic modulator (Gooch and Housego, AOMO 3080-125) or a mechanical chopper (Stanford Research Systems, SR542) was used to modulate the pump beam at 100 kHz or 1kHz. A mechanical linear motor stage (Newport, M-IMS600LM-S) was used to control the probe delay with respect to the pump. Both the pump and probe beams were focused onto the sample by an objective lens (Nikon, 10×, NA = 0.25). The transmitted probe beam was collimated through an aspheric lens (NA

= 0.6) and detected by an avalanche photodiode (APD; Thorlabs, APD430A/M). Spatial filters were used to optimize the profile of the beams. The change in the probe transmission induced by the pump was detected by a lock-in amplifier (HF2LI, Zurich Instruments). A two-axis Galvo mirror (Thorlabs GVS012/M) was used to scan the probe beam relative to the pump beam in space to image the carrier population in the sample.

**Data availability**

Source data are provided with this paper. All other relevant data are available from the corresponding authors on reasonable request.

**Code availability**

All calculations presented in this work are performed using publicly available standard packages. All relevant information used for reproducibility can be found in the text and Supplementary Information.

**Supplementary information**

Supplementary Figs. 1–18, Tables 1–4.

62    Carneiro, L. M. *et al.* Excitation-wavelength-dependent small polaron trapping of photoexcited carriers in α-Fe$_2$O$_3$. *Nat. Mater.* **16**, 819-825 (2017).

**Acknowledgements**

We acknowledge Dr. Baokun Liang for AC-HRTEM characterization. We acknowledge Dr. Jichao Zhang for XANES and EXAFS measurements. We acknowledge Dr. Ye Zou for XPS and UPS measurements. R.D. thanks National Natural Science Foundation of China (22272092; 22472085), Natural Science Foundation of Shandong Province (ZR2023JQ005), and Taishan Scholars Program of Shandong Province (tsqn201909047). P.P. is grateful to the European Union-NextGenerationEU, through the National Recovery and Resilience Plan of the Republic of Bulgaria, project No BG-RRP-2.004-0008 for the financial support and Discoverer PetaSC and EuroHPC JU for awarding access to DISCOVERER supercomputer resources. We acknowledge DESY (Hamburg, Germany), a member of the Helmholtz Association HGF, for the provision of experimental facilities. Parts of this research were carried out at PETRA III and we would like to thank Dr. Chen Shen for assistance in using beamline P08. Beamtime was allocated for proposal I-20230095. M.H. and S.C.B.M would like to thank Jonathan Perez for assistance with the GIWAXS measurements.

**Contributions**

S.F., T.Z., H.I.W., Z.W., R.D., X.F. and M.B. conceived the project. X.H. and S.F. contributed to the synthesis and routine characterization under the supervision of R.D., Z.W. and X.F. S.F. performed the TRTS measurements and analyzed the results under the supervision of H.I.W. and M.B. G.G. and K.L. conducted the TAM measurements and analyzed the data under the supervision of T.Z. P.P. contributed to the theoretical calculations. W.Z. and S.P. contributed to the variable-temperature conductivity measurements. U.K. coordinated the AC-HRTEM imaging and analysis. M.H. and S.M. contributed to the GIWAXS measurements. W.G. contributed to the STEM characterization. S.F., G.G. and X.H. wrote the manuscript with input from all authors. All authors discussed the results and commented on the manuscript.

**Corresponding authors**

Correspondence to T.Z., H.I.W, Z.W., R.D., X.F. or M.B.